\documentstyle[preprint,prd,aps]{revtex}

\begin{document}
\draft
\begin{titlepage}
\preprint{\vbox{\hbox{UDHEP-03-97}
\hbox{March 1997} }}
\title{ \large \bf Comments on Neutrino Tests of Special 
Relativity} 
\author{\bf S. L. Glashow$^{(a)}$, A. Halprin$^{(b)}$, 
P. I. Krastev$^{(c)}$, \\
C. N. Leung$^{(b), (d)}$ and J. Pantaleone$^{(e)}$}
\address{(a) Lyman Laboratory of Physics, 
Harvard University\\
Cambridge, Massachusetts 02138 \\} 
\address{(b) Department of Physics and Astronomy, 
University of Delaware\\
Newark, Delaware 19716 \\} 
\address{(c) School of Natural Sciences, The Institute for 
Advanced Study \\
Princeton, New Jersey 08540 \\}
\address{(d) Centro de F\'isica das Interac\c{c}\~oes 
Fundamentais \\
Instituto Superior T\'ecnico, Lisboa, Portugal\\}
\address{(e) Department of Physics and Astronomy, 
University of Alaska\\
Anchorage, Alaska 99508 \\}
\maketitle
\begin{abstract}

We point out that the assumption of Lorentz noninvariance 
examined recently by Coleman and Glashow leads to neutrino 
flavor oscillations which are phenomenologically equivalent
to those obtained by assuming the neutrinos violate the 
principle of equivalence.  We then comment on the limits 
on Lorentz noninvariance which can be derived from solar, 
atmospheric, and accelerator neutrino experiments.

\end{abstract}
\end{titlepage}

\newpage

In a recent paper\cite{CG}, Coleman and Glashow have proposed 
several interesting ways to test how well Lorentz invariance 
is obeyed in nature.  If Lorentz invariance is violated, one 
possible consequence is that the propagation of a free particle 
will depend on its identity.  In the case of massless neutrinos, 
this may lead to neutrino flavor oscillations because different 
neutrino species may have different maximum attainable velocities 
(which are no longer necessarily $c$).  For this to happen, it is 
necessary that the neutrino flavor eigenstates be distinct from 
their velocity eigenstates, defined to be the energy eigenstates 
at infinite momentum, so that a flavor eigenstate is a linear 
superposition of the velocity eigenstates and vice versa.  
If one considers the case of two neutrino mixing, say $\nu_e$ 
and $\nu_\mu$, the $\nu_e$ survival probability is given 
by\cite{CG}

\begin{equation}
P(\nu_e \rightarrow \nu_e) = 1 - \sin^2(2 \theta_v) 
\sin^2(\delta v E L/2),
\label{prob}
\end{equation}
where $\delta v = v_1 - v_2$ is the difference between the 
velocities of the velocity eigenstates $\nu_1$ and $\nu_2$, 
$\theta_v$ is the mixing angle:

\begin{equation}
\nu_e = \nu_1 \cos \theta_v - \nu_2 \sin \theta_v,~~~~~~
\nu_\mu = \nu_1 \sin \theta_v + \nu_2 \cos \theta_v,
\label{mix}
\end{equation}
$E$ is the neutrino energy and $L$ is the distance traveled 
by the neutrino.

We would like to point out that the energy dependence described 
in Eq.(\ref{prob}) is exactly the same as what one will get if 
one assumes that neutrinos violate the principle of equivalence 
in a certain way\cite{Gas},\cite{HL}.  This is interesting but 
not totally surprising because general coordinate invariance 
is violated in both cases.  The phenomenology of the case of 
equivalence principle violation has been studied in some 
detail over the last several years\cite{PHL} -\cite{MS}.  
The results of these studies can be straightforwardly 
translated to set limits on the possible violation of 
Lorentz invariance.  This is what we will discuss in the 
remainder of this paper.

We shall use the notation of Ref.\cite{HLP}.  It is easy to 
see that (see, e.g., Eqs. (14) to (16) in Ref.\cite{HLP}) 
the parameter $|\delta v|$, which measures the degree of 
violation of Lorentz invariance, should be compared with  
$2 |\phi \Delta \gamma|$ in Ref.\cite{HLP}, where $\phi$ is 
the gravitational potential in which the neutrino propagates 
and $\Delta \gamma$ is a parameter which measures the degree 
of violation of the equivalence principle.  The two are 
equivalent for the case of a constant gravitational potential.  
In addition, the mixing angle $\theta_v$ is equivalent to the 
mixing angle $\theta_G$ in Ref.\cite{HLP}.  Currently, there 
are positive indications of neutrino flavor mixing from solar 
and atmospheric neutrino experiments.  These data have been 
used to obtain allowed regions for the mixing parameters 
$\sin^2 (2 \theta_G)$ and $|\phi \Delta \gamma|$ (see 
Refs.\cite{PHL}, \cite{BKL} and \cite{HLP}), which can be 
translated directly into allowed regions for 
$\sin^2 (2 \theta_v)$ and $|\delta v|$.

First of all, due to the specific energy dependence, the 
solar neutrino data cannot be explained by long-wavelength 
vacuum oscillations caused by the Lorentz invariance 
violation.  This is because, if the mixing parameters are 
chosen such that enough $^8$B neutrinos are suppressed, 
there will not be sufficient suppression for the lower 
energy solar neutrinos, in contradiction to the data.  
It is therefore necessary to invoke the 
Mikheyev-Smirnov-Wolfenstein mechanism\cite{MSW} of 
matter enhanced transitions in the sun.  If we assume 
$\delta v$ to be constant inside the sun, the situation 
will be equivalent to the case of constant $\phi$ analyzed 
in Ref.\cite{HLP}.  Using the solar model of Bahcall and 
Pinsonneault\cite{BP}, one would find that a large portion 
of the mixing parameter space has already been excluded 
by the solar neutrino data, with two remaining allowed 
regions: a small mixing angle region for which 

\begin{equation}
|\delta v|~\sim~6 \times 10^{-19},~~~~~~0.002 < \sin^2 
(2 \theta_v) < 0.003, 
\label{smallangle}
\end{equation}
at 90\% confidence level, and a large mixing angle region 
for which 

\begin{equation}
4 \times 10^{-22} < |\delta v| < 4 \times 10^{-21},~~~~~~
0.38 < \sin^2 (2 \theta_v) < 0.81, 
\label{largeangle}
\end{equation}
also at 90\% confidence level.  Furthermore, the energy 
dependence implies that higher energy neutrinos have shorter 
oscillation length.  As a consequence, the higher energy 
atmospheric neutrino data imply a violation of Lorentz 
invariance in a small but overlapping parameter region (see 
Fig. 4 in Ref.\cite{HLP}).  It is quite remarkable that the 
mixing of two neutrinos is sufficient to account for both 
the solar neutrino and the atmospheric neutrino data. 

Aside from offering a possible resolution to the solar 
neutrino problem, velocity oscillations of neutrinos may 
provide the most sensitive tests of Lorentz invariance
and the equivalence principle.  Unlike conventional neutrino
oscillations, velocity oscillations become more important at 
higher energies.  Presently available accelerator data already 
provide useful constraints, as mentioned in Ref.\cite{CG} 
(see also Fig. 1 in Ref.\cite{HLP}).  In fact, part of the 
large angle region allowed by the solar neutrino data may be 
ruled out by the accelerator neutrino data.  Planned 
long-baseline neutrino oscillation experiments will be able 
to push the limit on $|\delta v|$ lower by one to two orders 
of magnitude, thereby limiting possible departures from 
special relativity or the equivalence principle. 

Whether neutrinos have observable masses is a central question 
of particle physics. It is often said that the observation of 
neutrino oscillations at accelerators (or their deduction from 
solar neutrino or cosmic ray experiments) would be conclusive 
evidence that at least one neutrino is massive. This is not true!  
Neutrino oscillations can also result from a tiny breakdown of 
Lorentz invariance and/or the principle of equivalence. More 
information than mere detection is needed to determine the 
underlying mechanism of neutrino oscillation.  The differing 
energy dependence between the mass mechanism and the 
mechanism due to Lorentz noninvariance (or due to equivalence 
principle violation) suggests that an accurate spectral 
measurement is required.  Super Kamiokande and SNO can
accurately measure the solar neutrino spectrum, and, as the 
analysis in Ref.\cite{BKL} shows, this measurement will test 
the viability of the small mixing region.  The current 
atmospheric neutrino data favor the large mixing region 
and this possibility will be tested by new atmospheric 
neutrino data from Super Kamiokande which should be available 
in the very near future.

It is of course a distinct possibility that neutrinos have 
nondegenerate masses and, at the same time, Lorentz 
invariance and/or the principle of equivalence is violated.
The phenomenology of neutrino oscillations in this case 
will be much more complicated.  As Coleman and Glashow 
pointed out, and as also discussed in section 2.3 of 
Ref.\cite{HLP}, for the case of two neutrino mixing there 
is an additional phase parameter beside the doubling of 
mixing parameters.  This will present a serious challenge to 
future neutrino experiments.

\newpage
\begin{center} 
{\bf ACKNOWLEDGEMENTS}\\
\end{center}

This work was supported in part by the U.S. Department of Energy 
under Grant No. DE-FG02-84ER40163 and by the National Science 
Foundation under Grant No. NSF-PHYS-92-18167.  The work of P.K. 
was partially supported by NSF grant \#PHY-9513835.  C.N.L. would like 
to thank G. C. Branco for bringing Ref.\cite{CG} to his attention.  
He is also grateful for the hospitality extended to him by G. C. 
Branco, L. Ferreira, J. Pulido, M. Rebelo, E. Ribeiro, A. Rossi, 
V. Vieira and the other members of the Centro de F\'isica das 
Interac\c{c}\~oes Fundamentais at the Instituto Superior T\'ecnico, 
where part of this work was carried out. 
\raggedbottom

\end{document}